# Transverse azimuthal dephasing of vortex spin wave in a hot atomic gas

Shuai Shi[1,2], Dong-Sheng Ding[1,2,†], Wei Zhang[1,2], Zhi-Yuan Zhou[1,2], Ming-Xin Dong[1,2], Shi-Long Liu[1,2], Bao-Sen Shi[1,2,*] and Guang-Can Guo[1,2]

[1]*Key Laboratory of Quantum Information, University of Science and Technology of China, Hefei, Anhui 230026, China*

[2]*Synergetic Innovation Center of Quantum Information & Quantum Physics, University of Science and Technology of China, Hefei, Anhui 230026, China*

*Corresponding author:* [†]*dds@ustc.edu.cn,* [*]*drshi@ustc.edu.cn*

Optical fields with orbital angular momentum (OAM) interact with medium have many remarkable properties with its' unique azimuthal phase, showing many potential applications in high-capacity information processing, high precision measurement etc. The dephasing mechanics of optical fields with OAM in an interface between light and matter plays a vital role in many areas of physics. In this work, we study the transverse azimuthal dephasing of OAM spin wave in a hot atomic gas via OAM storage. The transverse azimuthal phase difference between the control and probe beams is mapped onto the spin wave, which essentially results in dephasing of atomic spin wave. The dephasing of OAM spin wave can be controlled by the parameters of OAM's topological charge and beam waist. Our results are helpful for studying OAM light interaction with matter, maybe hold a promise in OAM-based quantum information processing.

Studying and controlling of the quantum states of the collective excitations of atoms is conducive to the development of quantum information technology [1, 2]. Specially, coherent storage and manipulation of quantum superposition states is essentially important to ensure high-fidelity evolution. Recently, quantum state with atomic decoherence attracts many researchers in quantum optics and atomic physics [3, 4], which shows a reliable and long-lived storage units for quantum communications [5].

Light pulse can be stored as collective excitations and read out in an atomic vapor [6, 7].

Even though, the transverse amplitude and phase profile of the pulse can be preserved very well for a short time storage [8], but the fidelity of storage for a long time will be seriously affected by the atomic motion during the storage [9]. There are two main kind dynamics underlie the decoherence mechanisms in the atomic vapor: the collisions among atoms and with the internal wall of the vapor cell damage the internal state of the atoms, and the random thermal motion of the atoms [3]. The decoherence caused by collisions can be improved effectively by coating the inner walls of a cell with an antirelaxation film, and adding buffer gas into the cell to keep the atoms in the illuminated region for a longer time [10, 11]. Although the atomic motion can be utilized to freeze a light pulse in a coherently driven atomic medium and implement controllable slow light beam splitter [12, 13], but it crucially affects the resolution and coherence time of the stored optical images. The decoherence mechanism induced by atomic motion attracting a lot of research, thus a lot of technology used to overcome and reduce the decoherence are proposed, such as: atomic motion induced diffraction can eliminate the paraxial diffraction at proper two-photon detuning [14], Spin echo technology can be used to extend the atomic coherence time [15, 16], storing the Fourier transform of the image can overcome the adverse effects of diffusion [17], etc.

Light carrying OAM stored in hot atomic gas can maintain its phase singularity due to topological stability [20]. Storing light carrying OAM in hot atomic gas not only is helpful for high-capacity quantum information processing [21, 22], but also can exploit the advantages of warm-vapor-cell system in relative simplicity; easy controlling of atomic density etc. However, various decoherence mechanisms significantly influence the fidelity of the collective excitations of atoms, such as random dephasing, loss of atoms and atomic motion etc [4]. As a result, studying the dephasing of the OAM light with thermal motion of hot atomic gas is crucially important. The phase front fluctuation induced from atomic random motion in the vector direction of the SW [23] directly affects the fidelity. While in the transverse radial direction, atoms from all directions carrying a phase destructively interfere at the dark center which is a phase singularity of the SW [20, 24]. The dephasing in the transverse azimuthal direction resulting from a phase front fluctuation has not been observed before, which is the topic of this work.

In this work, we introduce different OAM to the control beam to prepare SW with different topological charge. We show that, SW with higher topological charge experience larger decoherence, and a theoretical analysis for the distinct decoherence mechanisms is given. Theoretical model based on the random motion of atoms predicts that the decoherence is also related to the beam waist, which is in good agreement with experimental observation.

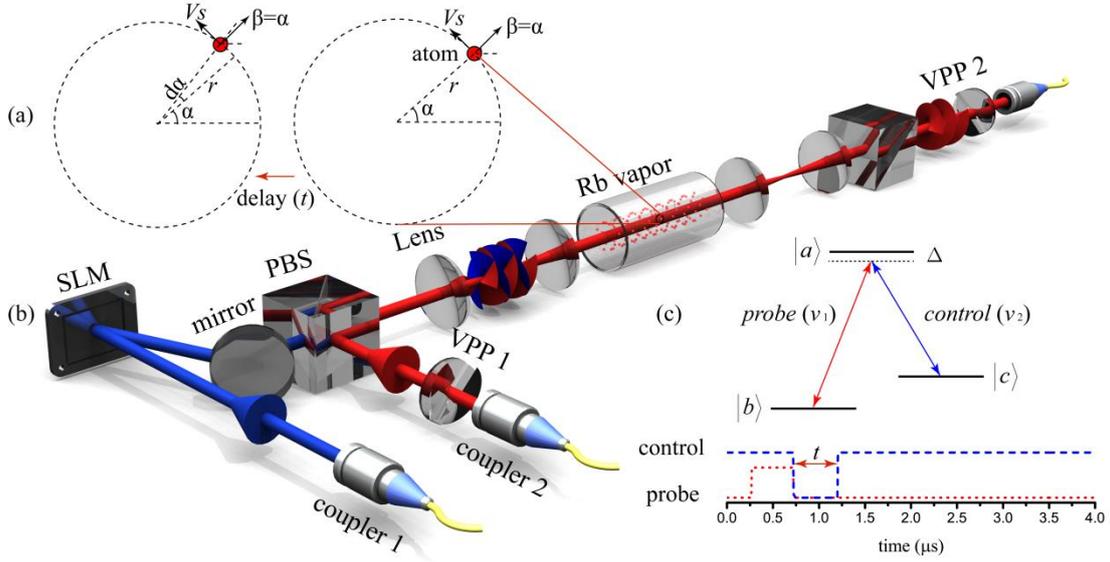

Fig .1 Experimental setup for optical vortex storing. (a) Illustration of the SW dephasing induced by atomic random motion along the azimuthal direction. (b) The experimental setup. SLM-spatial light modulator, VPP-vortex phase plate, PMT- photomultiplier tube, PBS-polarizing beam splitter. (c) The energy level scheme of the D1 transition of $^{85}$Rb showing the three levels of the $\Lambda$ system. |a>, |b> and |c> correspond to atomic states $5^2P_{1/2}$ F=2, $5^2S_{1/2}$ F=2 and $5^2S_{1/2}$ F=3 respectively.

## Results

The storage experiments are performed within the D1 transition of $^{85}$Rb. The energy level scheme is presented in Fig.1 (c), showing the $\Lambda$ system and the pump and probe transitions. An external cavity diode laser is stabilized to the F=3→F'=2 transition. The laser is divided into two beams of perpendicular linear polarizations, the pump and the probe. The pump beam is passing through an acousto-optic modulator (AOM), allowing us to control both the frequency and the intensity of it. The pump is detuned by 300 MHz to the red of the F=3→F'=2 transition, and the probe is set 3.035 GHz to the blue of the pump after double-passes through an AOM. A schematic of the experimental setup is shown in Fig. 1(b). The control beam from coupler 1 is diffracted off a computer-generated fork-diffraction pattern on the spatial light modulator (SLM; Holoeye LETO

LCoS). The fork dislocation in the patterns introduces helical phase fronts (exp(i$m$α)) to the first-order diffracted pulse, α representing an azimuthal angle, thereby imparting an OAM of $m$. The probe beam from coupler 2 gets OAM $n$ after it passing through the vortex phase plate 1 (VPP). The control and the probe are recombined on a polarizing beam splitter (PBS) and copropagate toward the vapor cell. The plane of the SLM and the VPP 1 are imaged onto the center of the vapor cell by using a 4$f$ imaging system, which consists of two lenses of focal length $f$=300mm. So the control and the probe are shaped as Laguerre-Gaussian (L-G) beams with a waist of $w$=2mm (at the center of the vapor cell). The total intensity of the control is 9 mW. A 5 cm long vapor cell containing $^{85}$Rb is used. The temperature of the cell is stabled at 55 ℃, providing a rubidium vapor density of ~2.2×10$^{11}$/cc. The cell is placed inside a five-layered magnetic shield. After the beams pass through the vapor cell, another PBS is used to isolate the control beam, and a 4$f$ imaging system is used to image the plane of the VPP 1 onto the VPP 2. VPP2 with a helical surface opposite to VPP 1 can be used to "flatten" the phase of the probe beam, and then the probe beam is collected into coupler 3. In addition to polarization filtering, we also performed a frequency filtering by using a temperature-controlled Fabry-Perot (F-P) etalon. Finally, the probe beam is detected by a photomultiplier tube (PMT HAMAMATSU H10721-01). The frequency shift between the control and the probe is setting to the center of the EIT resonance.

A typical experimental sequence started by applying the control beam for a long duration, optically pumping a substantial atomic population to the $|F=2\rangle$ state. A vortex probe pulse with duration 500 ns is then sent into the cell. The control beam is turned off to store the probe pulse as the atomic ground-state coherence when the probe pulse is propagating in the vapor cell. After a certain storage duration (during which decoherence of the atoms occurred), the control beam is turned back on, retrieving the probe pulse, which is finally detected by the PMT. The effect of decoherence on the retrieved probe pulse is studied by measuring it for different storage durations.

In the main experiment, VPP 1 introduces OAM $n$=2 to the probe pulse, we study the effect of dephasing by using control beam with OAM $m$= 2, 0 and -2. The measured lifetime τ decreases as the OAM difference between the control and the probe increases, which implies that the decoherence is related to the azimuthal phase gradient of the stored spin wave (SW).

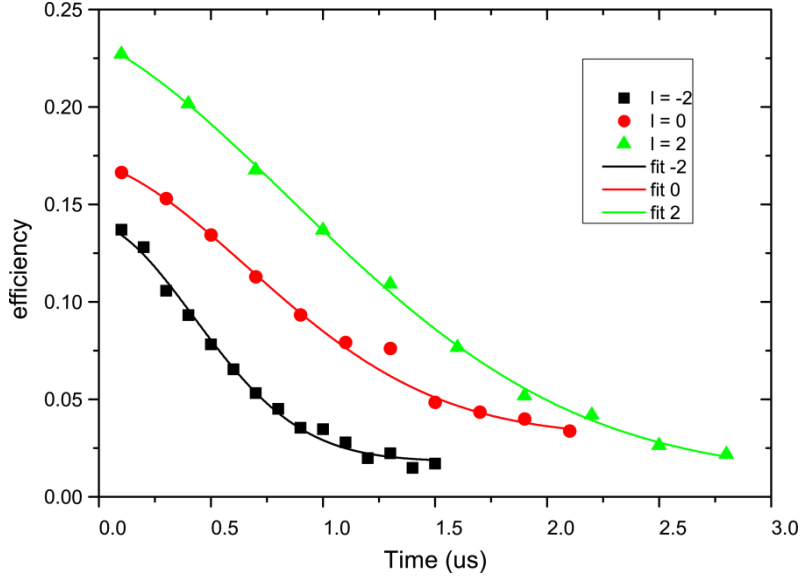

Fig. 2 Lifetime measurement results for control beam with different OAM, the probe beam with OAM 2. The data are fitted by using $\eta(t) = C_1 + C_2 e^{-t^2/\tau_D^2} e^{-t^2/\tau_0^2} e^{-t/\tau_1}$. The black curve is control beam with OAM -2, lifetime $\tau_D = 0.74 \mu s$; The red curve is control beam with OAM 0, lifetime $\tau_D = 1.6 \mu s$; The green curve is control beam with OAM 2, no azimuthal dephasing. All of these curves have a common longitudinal lifetime $\tau_0 = 1.81 \mu s$ and a lifetime $\tau_1 = 3.78 \mu s$ due to other decoherence mechanisms.

The experimental result is shown in the Fig. 2. We find that the decoherence mechanism of the storage can be explained by the dephasing of the SW induced by atomic random motion. The longitudinal decoherence mechanism has been explored in previous experiments [16, 23], but the transverse decoherence mechanism has not attracted sufficient attention yet. As the longitudinal decoherence is related to the wavelength of the SW, the transverse decoherence is related to the topological charge of the SW, which is determined by the OAM difference between the control and the probe.

We assume that the Rabi frequency of the control and the probe is $\Omega_1$ and $\Omega_2$, respectively. The interaction between light and atoms is governed by the Hamiltonian:

$$H_I = -\frac{\hbar}{2}\Omega_1 e^{im\alpha}|a\rangle\langle b| - \frac{\hbar}{2}\Omega_2 e^{-i\phi_2} e^{in\alpha}|a\rangle\langle c| + H.c. \qquad (1)$$

here exp(i$l\alpha$) is azimuthal phase, which reflects the OAM of the control and the probe. We use collective, slowly varying atomic operators describe the quantum properties of the atoms.

$$\hat{\sigma}_{\alpha\beta}(z,t) = \frac{1}{N_z}\sum_{j=1}^{N_z}|\alpha_j\rangle\langle\beta_j|e^{-i\omega_{\alpha\beta}t} \qquad (2)$$

Here $N_z$ is the particles number contained in the volumes at position z.

The atomic evolution is governed by a set of Heisenberg-Langevin equations [6]. Under the assumption that the Rabi frequency of the probe is much smaller than the control and that the number of photons in the input pulse is much less than the number of atoms. Thus the lowest nonvanishing order of $\hat{\sigma}_{bc}(z,t)$ is:

$$\hat{\sigma}_{bc}(z,t) = -g\frac{\Omega_2}{\Omega_1}e^{i(n-m)\alpha} \qquad (3)$$

We can decelerate and stop the input light pulse by adiabatically turn off the control beam. In this process, the azimuthal phase difference between the probe and the control is mapped onto collective states of matter in which they are stored.

The decoherence induced by atomic random motion can be divided into three kinds, as atoms moving in three dimensional space. The first one is caused by the atoms random motion along the wavevector direction of the SW, resulting in a phase fluctuation [23]. The second one is movement along the radial direction, atoms from all directions carrying a phase destructively interfere at the dark center which is a phase singularity of the SW [20, 24]. The final one is movement along the azimuthal direction, resulting in a phase front fluctuation which can be understood as follows. We will introduce the $l$=n-m=1 case for intuitively understood. As shown in Fig. 1(a) an optical pulse with OAM is stored in the atomic ensemble as SW with azimuthal phase β equal to the azimuthal angle α, and will be retrieved after a time delay *t*. During this interval, each atom with phase β moves from one azimuthal point to another randomly. The internal states of the atoms are conserved, however, the azimuthal motion of the atom leads to a perturbation on the phase front of the SW. Consequently, the projection of the perturbed SW on the original state gradually decreases as the delay of the retrieval is increased. Therefore, the atomic azimuthal motion leads to a random phase front fluctuation to the SW and thus causes decoherence. The timescale of the dephasing can be estimated by calculating the average time for the atoms needed to cross 1/2π of the azimuthal period of the SW. Since atoms at different radial position need to cross different distance to move the same azimuthal angle, the lifetime $\tau_D(r)$ is changing with radial position r: $\tau_D(r) \sim (r/l\nu_S)$, with $\nu_S = \sqrt{k_B T/m}$ the one-dimensional average speed, where $k_B$ is the Boltzman constant, T the average temperature of the atoms, and 2π/$l$ the azimuthal period angle of the SW. The intensity of the spin wave is in Gaussian distribution due to the 4*f* system imaging the plane of the SLM and the VPP 1 onto the center of the vapor cell. The Gaussian weighted average lifetime is:

$$\tau_D \sim \int \tau_D(r) \frac{4r}{W_0^2} \exp[-\frac{2r^2}{W_0^2}]dr = \frac{\sqrt{2}\pi W_0}{4l\nu_S} \quad (4)$$

Here $W_0$ is the beam waist. A more detailed calculation yields that the retrieval efficiency: $\gamma(t) \sim e^{-t^2/\tau_D^2}$, with a lifetime $\tau_D = \frac{\sqrt{2}\pi W_0}{4l\nu_S}$. Assume the *j*th atom is excited to $|\psi_{j0}\rangle = e^{il\alpha_j(0)}|c\rangle$ at time t=0, and moves to azimuthal position $\alpha_j(t) = \alpha_j(0) + (v_j/r)t$ after a storage time of t. The state freely evolves to $|\psi_{jt}\rangle = e^{il\alpha_j(t)}|c\rangle$, the retrieval efficiency of the *j*th atom is proportional to the overlap between the original state and the perturbed one

$$\gamma_j(t) \sim |\langle\psi_{j0}|\psi_{jt}\rangle|^2 = |e^{i(v_j/r)t}|^2 = |\int f(v)e^{i(v/r)t}dv|^2 \quad (5)$$

With $f(v) \sim e^{-mv^2/2k_BT}$ is a Boltzmann distribution of the velocity at temperature T. Integrating over all possible velocities, we obtain $\gamma_j(t) \sim e^{-t^2/\tau_D^2}$, with the lifetime $\tau_D(r) = (r/l\nu_S)$. This atom contribute 1/n of the overall retrieval efficiency, the total retrieval efficiency is $\gamma(t) = \frac{1}{n}\sum_j \gamma_j(t)$, with n the number of total excited atoms. The Gaussian weighted average retrieval efficiency is: $\gamma(t) \sim e^{-t^2/\tau_D^2}$, with the lifetime $\tau_D = \frac{\sqrt{2}\pi W_0}{4l\nu_S}$. In our case, the OAM difference between the control and the probe determines the topological charge *l*=n-m, which related to the lifetime of the SW. In order to confirm the decoherence mainly caused by atomic azimuthal motion, we increase the topological charge of the SW by decreasing the OAM of the control (see Fig. 2). According to the above model, the dephasing is enhanced and the lifetime will be shortened. In our experiment, we use the control with OAM 2, 0 and -2 and measure the lifetime of the quantum memory for each case. The Experimental results are shown in Fig. 2. Fitting the data in Fig. 2 with total decay function:

$$\eta(t) = C_1 + C_2 e^{-t^2/\tau_D^2} e^{-t^2/\tau_0^2} e^{-t/\tau_1} \quad (6)$$

Here $C_1$ is the background noise level and $C_2$ is the retrieval efficiency at t=0, the decay includes three parts: The first one $e^{-t^2/\tau_D^2}$ has been explained above, the second one $e^{-t^2/\tau_0^2}$ is caused by atomic longitudinal random motion [23], and the third one $e^{-t/\tau_1}$ is caused by other decoherence processes [4]. We obtain lifetimes of $\tau_D = 1.6\mu s$ for $l$=2 and $\tau_D = 0.74\mu s$ for $l$=4, moreover, all of these curves have a common longitudinal lifetime $\tau_0 = 1.81\mu s$ and a lifetime $\tau_1 = 3.78\mu s$, that is because the wavelength and other properties of the SWs are the same. As expected, the transverse dephasing of the SW dominates in this process. The lifetime is decreasing as the topological charge of the SW is increasing. Our results clearly show that the dephasing of the SW is sensitive to the OAM difference between the probe and the control, and that SW with high topological charge is extremely sensitive to the atomic random motion. The above model is in good agreement with the experimental results.

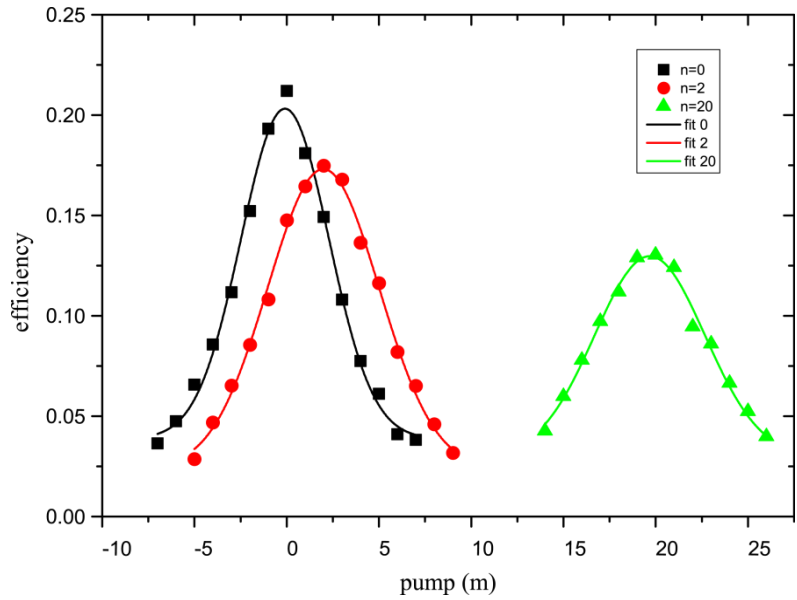

Fig. 3 Retrieval efficiency curves along with the change of the OAM m of the control. The data are fitted by using $\eta(m) = C_1 + C_2 e^{-B(m-n)^2}$. The black, red and green curves corresponding to probe with OAM $n$= 0, 2 and 20 respectively.

To further confirm the relationship between the retrieval efficiency and the topological charge of the SW, we measure the retrieval efficiencies after storage of 0.5 μs for SWs with different topological charge (Fig. 3). We found that the dependencies follow Gaussian curves with the center corresponding to SW with zero topological charge.

This relationship can be understood as follows. We take the azimuthal lifetime $\tau_D = \frac{\sqrt{2}\pi W_0}{4l\nu_S}$ into the total decay function (6), and assume that all parameters except *l* are constant, then we get: $\eta(l) = C_1 + C_2 e^{-Bl^2}$, it is clearly a Gaussian function. We further verify that this is valid for probe with different OAM 0, 2 and 20.

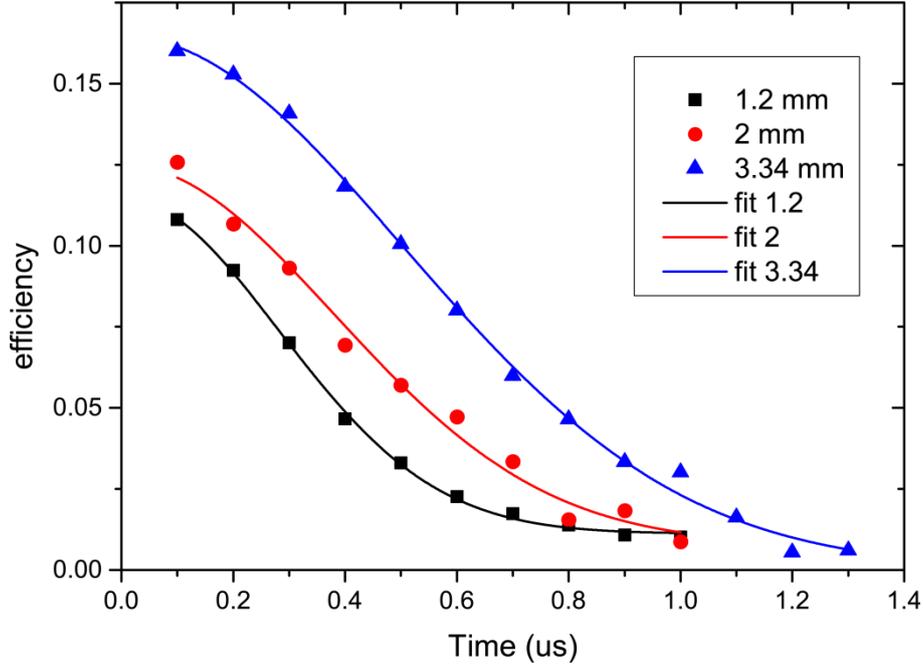

Fig. 4 Lifetime measurement results for SWs with different waists. The black curve is for the waist 1.2mm, lifetime $\tau_D = 0.427\mu s$; the red curve is for the waist 2 mm, lifetime $\tau_D = 0.61\mu s$; the blue curve is for the waist 3.34mm, lifetime $\tau_D = 0.88\mu s$

Finally, we prove that the lifetime is also related to the waist of the SW, which is determined by the waist of the probe and control beams. We change one of the two lenses in the 4*f* system to zoom in and out the wait of the beams. As expected, the lifetime increases from 0.427 to 0.883 μs when increasing the waist of the SW from 1.2 to 3.34 mm. (See Fig. 4)

## Discussion

In our experiment, we have divided the decoherence mechanisms induced by atoms random motion into three kinds. And we thoroughly investigated the transverse azimuthal dephasing of the stored SW by varying its topological charge. A theoretical explanation in accordance with experimental results is given. According to the theory, the decoherence of SW with topological charge is induced by the transverse motion of atoms, so experimental schemes which can eliminate the transverse motion of atoms are suitable for high dimensional light storage, such as: optical lattice, Rb-filled photonic crystal fiber etc.

## Acknowledgments

This work was supported by the National Fundamental Research Program of China (Grant No. 2011CBA00200), the National Natural Science Foundation of China (Grant Nos. 11174271, 61275115, 61435011, 61525504).